# Differential Conductance Measurements of $MgB_2$-Based Josephson Junctions Below 1 Kelvin

Steve Carabello, Joseph Lambert, Jerome Mlack, and Roberto Ramos

***Abstract*—Magnesium diboride has many intriguing characteristics, including its relatively high critical temperature and two-band nature. Most prior studies of $MgB_2$ thin film Josephson junctions have been conducted above 2 Kelvin. We report results of sub-1 Kelvin experiments of $MgB_2$/insulator/Pb junctions whose *a-b* plane is exposed for electron tunneling. By measuring differential conductance at low temperature, new details in the structure of the sigma- and pi-band gaps are observed in this data, consistent with theoretical predictions.**

## I. Introduction

SOON after the discovery of superconductivity in $MgB_2$, the existence of two superconducting energy gaps was clearly shown by a wide variety of techniques, including angle-resolved photoemission spectroscopy (ARPES), specific heat, point-contact spectroscopy, Raman spectroscopy, photoemission spectroscopy, and scanning tunneling spectroscopy [1]. These two distinct gaps result from different pairing strengths between phonon modes and the $\sigma$ bonds localized in the boron planes (which have a strong coupling) and the $\pi$ bonds perpendicular to the planes (which have a weak coupling) [2].

A variety of calculations have used the electron-phonon coupling to derive the values of these energy gaps from first principles [2-8]. In the analysis by Choi et. al. [2,4], the Eliashberg equation was solved in the clean limit, taking into account the fully anisotropic electron-phonon interaction and the anharmonicity of the phonons. This analysis resulted in a good prediction of the critical temperature, as well as clear evidence of two energy gaps. The resulting density of states as a function of energy gap at 4K is given in Fig. 1 [2]. As expected, those modes associated with the $\sigma$ bonds give rise to strong Cooper pairs with a large energy gap (between 6.4 and 7.2 meV), while those associated with the $\pi$ bonds give rise to a separate, smaller energy gap (between 1.2 and 3.7 meV).

Of particular interest is the substructure evident within each of these two peaks, which have generated some controversy. Reference [9] suggests, among other objections, that this "distribution of gaps *within* the $\sigma$ and the $\pi$ sheets" should not be observable in real samples, due to the need for unreasonably small scattering rates. However, very high quality thin films have been found to meet or exceed the purity of single crystals [10, 11]. Indeed, recent tunneling spectroscopy experiments incorporating such films at temperatures from 7.0 to 1.8K have exhibited energy gap substructures consistent with these theoretical predictions [12].

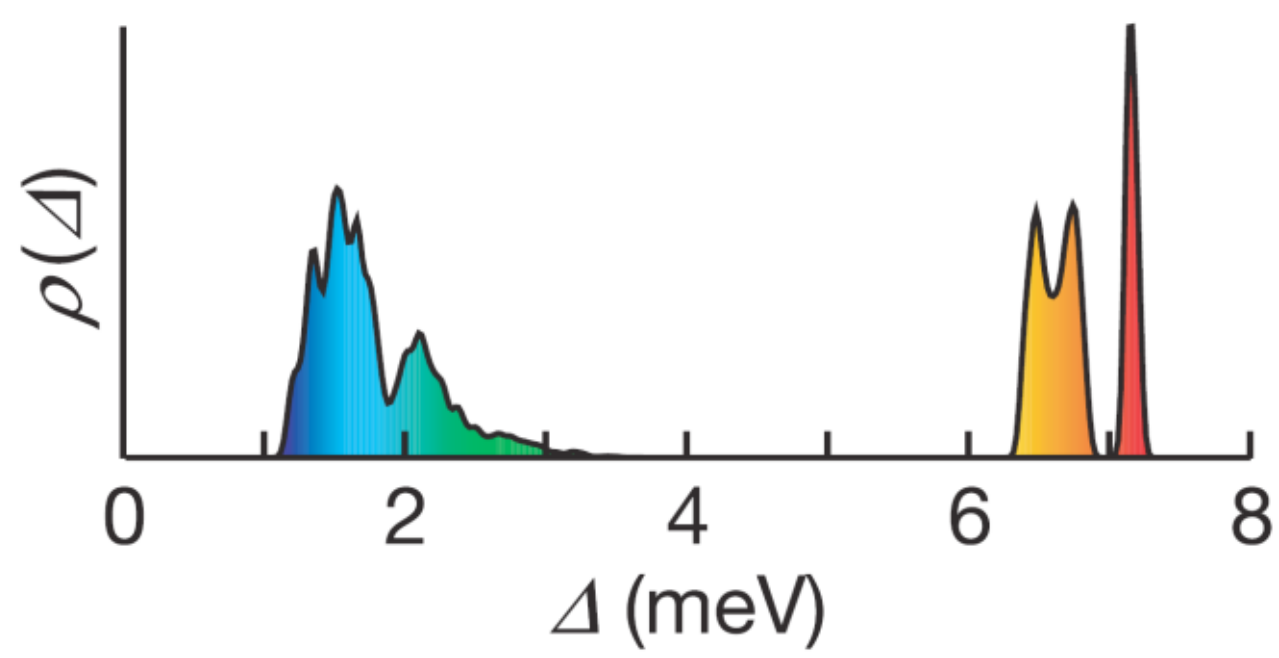

Fig. 1. Theoretical predictions of substructure in the energy gaps of $MgB_2$ [2]. $\rho(\Delta)$ represents the density of states as a function of energy gap, calculated by integrating the "local gap distribution" (introduced in [2, 4]).

We have explored the superconducting energy gaps of $MgB_2$ using electron tunneling spectroscopy of Josephson junctions below 1K. The differential conductance curve of a tunnel junction has a peak at the sum of the energy gap values of the two leads $(\Delta_1 + \Delta_2)/e$, and so provides a direct measure of the energy gap [13]. The $MgB_2$/native oxide/Pb tunnel junction used in this study incorporates a high purity $MgB_2$ thin film grown by hybrid physical-chemical vapor deposition (HPCVD) on a single crystal 6H-SiC substrate [14]. As this film formed, the *c* axis of the $MgB_2$ was tilted away from film normal, exposing the *a-b* plane for tunneling. As a result, tunneling occurred both along the *c*-axis, and along the boron planes, providing access to both the $\sigma$ and the $\pi$ gaps [14].

## II. Our Experiment

The experiments were performed in a helium dilution refrigerator with a base temperature of 20 mK. Current and voltage lines were filtered using thermocoax cables and LC filters encased in copper powder matrix. All leads were thermally grounded at the 1K pot, still, and mixing chamber.

The key data for this experiment are derived by gathering current vs. voltage (*I-V*) curves. The current bias for our

Manuscript received 3 August 2010. This work was supported in part by a grant in aid by Sigma Xi, for S. Carabello.

S. Carabello, J. Lambert, J. Mlack and R. Ramos are with the Department of Physics, Drexel University, Philadelphia, PA 19104 USA (S. Carabello: phone: 215-895-1759; e-mail: sac69@drexel.edu).

junction was provided by an Agilent 3220A function generator, swept at 50mHz, passing through a bias resistor. The voltage across the junction was amplified, then recorded at a sampling rate of 10kHz by a NI 9215 16-bit data acquisition card. From the *I-V* curves thus gathered, differential conductance (*dI/dV*) was derived numerically.

## III. Noise Reduction: Averaging Techniques

Despite the extensive filtering, low-frequency noise remained in our data. Given the extremely high density of points due to the high sample rate, even tiny fluctuations in the signal were translated into large discrepancies in the *dI/dV-V* curve. As a result, a series of averaging steps was necessary to produce useful results.

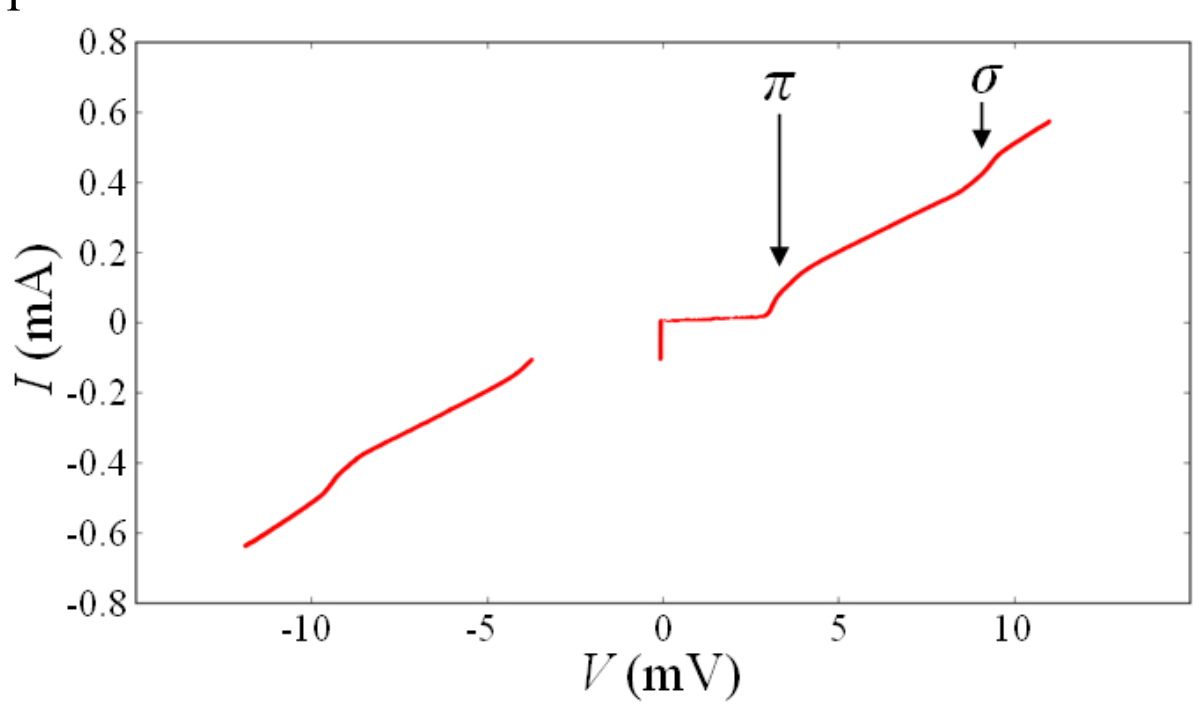

Fig. 2. Raw *I-V* data for a single sweep (from high current to low current) of our $MgB_2$/I/Pb junction. Changes in slope corresponding to both the $\sigma$ and $\pi$ bands are clearly visible, as is the hysteretic nature of our junction below 1K.

Fig. 2 shows a typical *I-V* curve of a single sweep (from high current to low current) in this experiment. Hysteretic behavior is clearly visible, as are the changes in slope corresponding to the $\pi$ and $\sigma$ energy gaps. On this scale, the curve appears quite clean, with little evidence of noise.

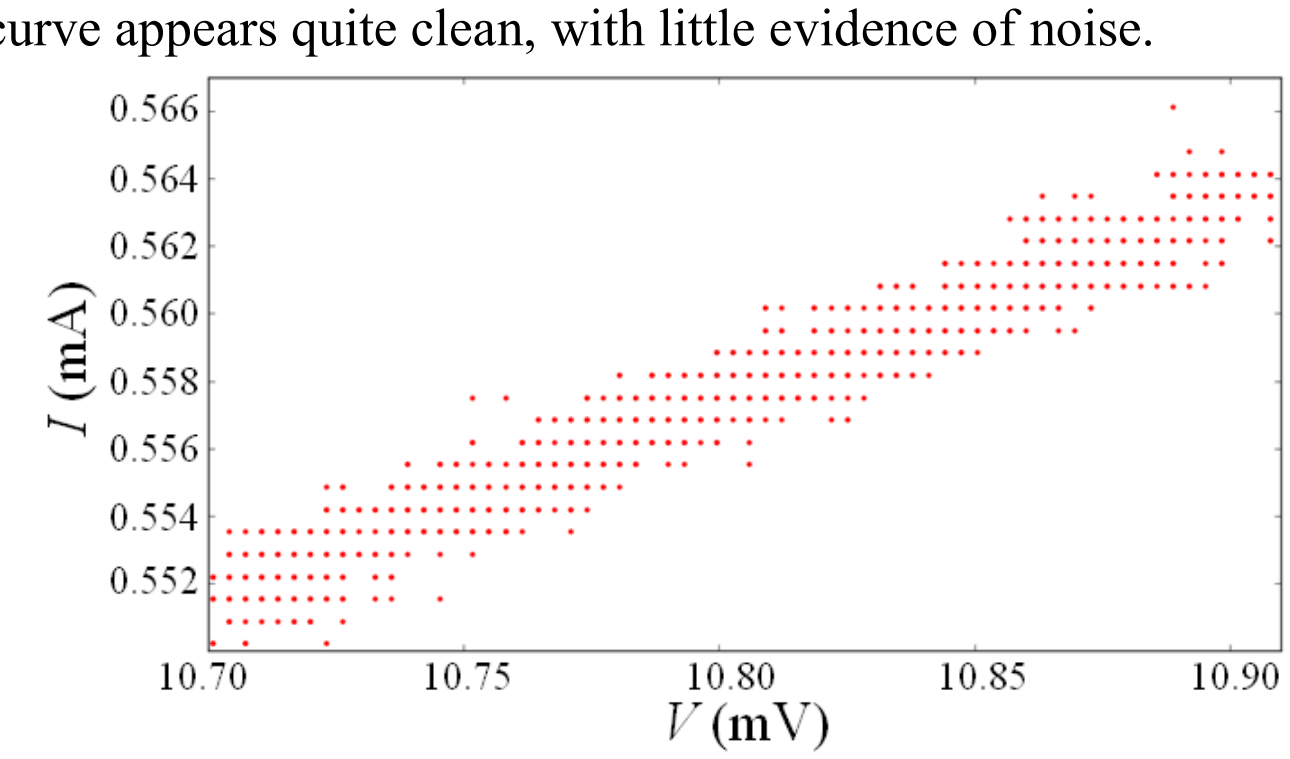

Fig. 3. A small segment of the raw *I-V* data for a single sweep of our $MgB_2$/I/Pb junction. Note the discrete nature of this digital data.

Zooming in, one can see the discrete nature of this digital data (Fig. 3). As a result, the raw data points cannot form useful conductance curves directly. If these points were randomly distributed, then averaging even a small number would produce a smoother curve. Fig. 4 shows the same data after computing a running average of 20 data points. Periodic structures, with a period of 167 data points (corresponding to a frequency of 60 Hz) are clearly evident. So, all data used for computing conductance curves were first averaged over 167 data points, to minimize the effect of this spurious noise. (See Fig. 5.)

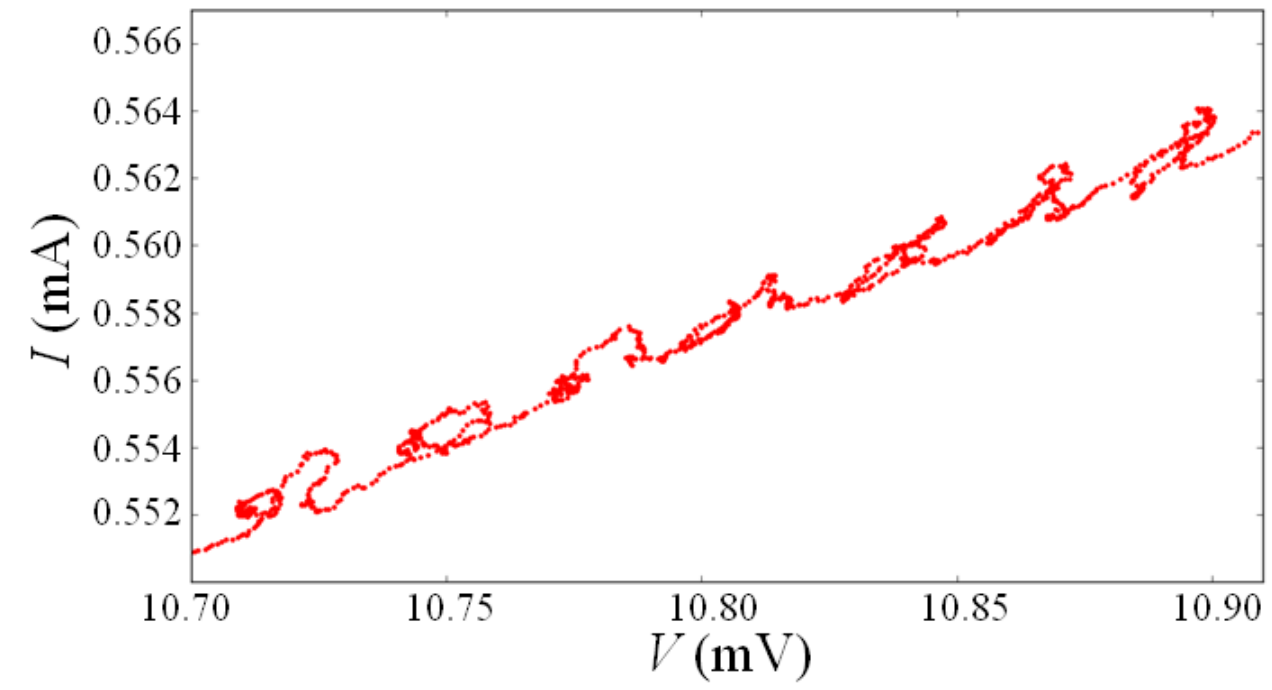

Fig. 4. The same data as in Fig. 3, after performing a running average over 20 data points. The periodic structures are evidence of 60 Hz noise.

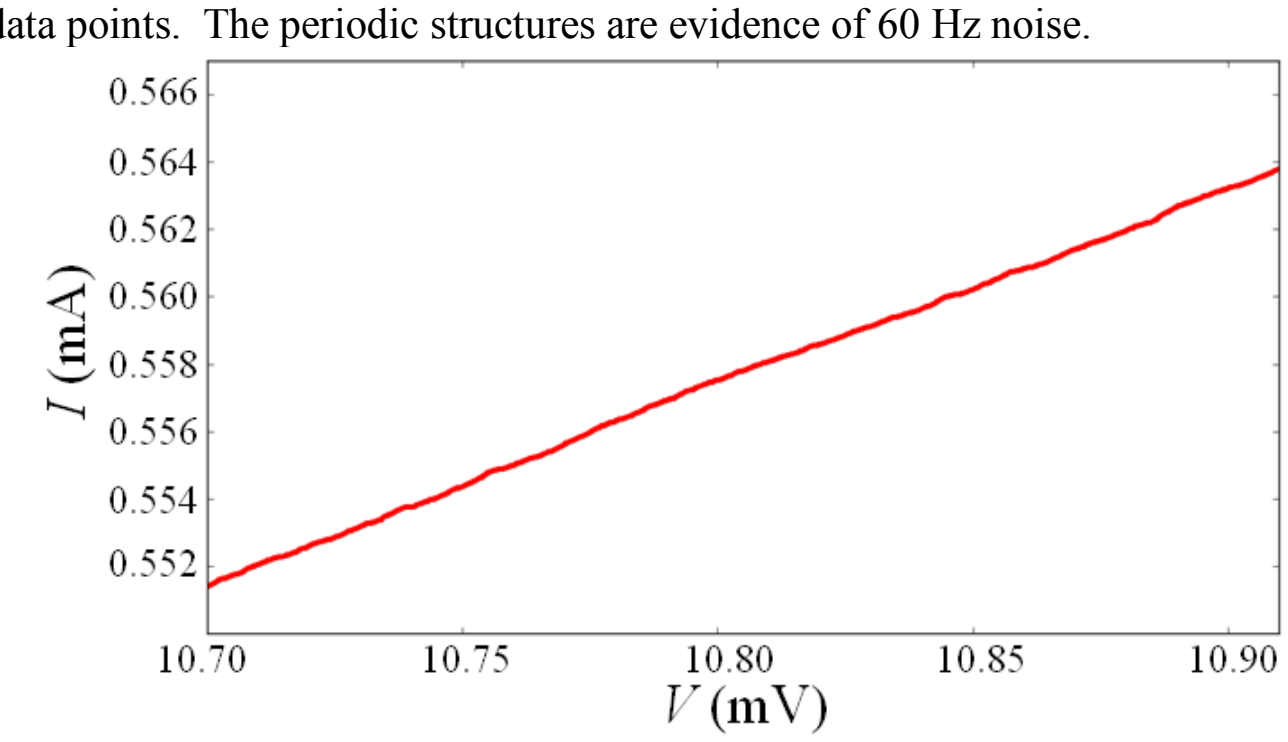

Fig. 5. The same data as in Fig. 3, after performing a running average over 167 data points. The 60 Hz noise has been nearly eliminated.

As a final step prior to finding a conductance curve, we binned this averaged *I-V* data into separate data points approximately 0.03mV apart. This allows us to take advantage of multiple sweeps through the *I-V* curve, thereby causing each data point in our results (Fig. 6) to be the result of averaging thousands of raw data points. This final averaging also serves to provide additional noise suppression. It is important to note, however, that all of the main features in the conductance curves persist while using a wide variety of averaging techniques, and are even apparent (though with less clarity) with little averaging at all.

## IV. Our Results

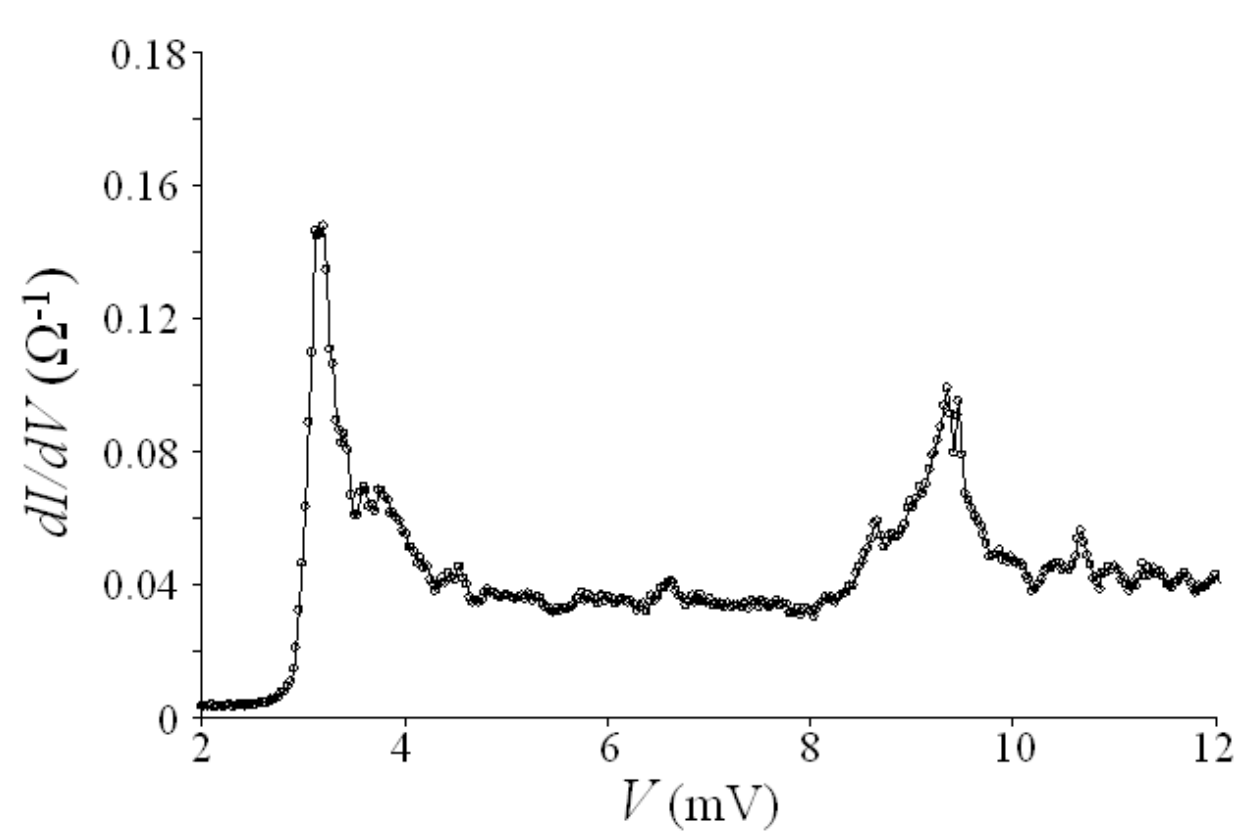

Fig. 6. Differential conductance curve at *T*=350mK. As anticipated by the theory of Choi et al. [2], evidence is seen for substructure within each energy gap. Additionally, the number of features, and their approximate voltages, correlate well with theory.

The resulting differential conductance curves exhibit features very similar to theory; Fig. 6 shows the right branch of such a curve, using data gathered at $T$=350mK. Corresponding to theory, there are several prominent features; each peak corresponds to $(\Delta_{MgB} + \Delta_{Pb})/e$, with $\Delta_{Pb}$ = 1.376 meV at the temperatures of our experiment.

These results, and those on a similar thin film device measured from 7.0-1.8K [12], are the first to clearly illustrate the substructure within each of the energy gaps of $MgB_2$ in this way. In contrast to the theory (Fig. 1), the $\pi$ peak is taller than the $\sigma$ peak. This result is expected, since the Pb electrodes of our junction had a large percentage of its contact with the *c* axis, and less along the *a-b* plane.

The tallest point on the differential conductance curve (corresponding to the $\pi$ peak), occurs at a voltage of 3.2 mV. Subtracting $\Delta_{Pb}$, we find an energy gap value of 1.8 meV. The highest voltage energy gap peak (corresponding to the $\sigma$ peak), occurs at a voltage of 9.3 mV. Again subtracting $\Delta_{Pb}$, we get an energy gap value of 7.9 meV. These results are reasonably consistent with theory. One less prominent peak to the right of the main $\pi$ peak, and substructure to the left of the $\sigma$ peak, are also consistent with theory.

## V. Conclusion

We have observed substructure in the energy gaps of $MgB_2$ using $MgB_2$/I/Pb Josephson junctions, via differential conductance measurements. These results are consistent with theoretical predictions using the anisotropic Eliashberg formalism, and were made possible due to the low scattering rates in high-quality $MgB_2$ thin films. By displaying such fine details, our data helps to illustrate the importance of the anharmonicity of the phonon modes in $MgB_2$ in making theoretical predictions.

## Acknowledgment

We thank X. X. Xi and Ke Chen (Temple University, Philadelphia, PA USA) for providing high-quality $MgB_2$-based junctions, and many important suggestions and useful insights.